\documentclass[aps,prb,twocolumn,superscriptaddress,showpacs]{revtex4}

\usepackage{amsmath}
\usepackage{amssymb}
\usepackage{graphicx}
\usepackage{pstricks}
\usepackage{scalefnt}
\usepackage{tikz}
\usepackage{color}
\definecolor{Blue}{rgb}{0.3,0.3,1}
\newcommand{\be}{\begin{equation}}
\newcommand{\ee}{\end{equation}}
\newcommand{\bpm}{\begin{pmatrix}}
\newcommand{\epm}{\end{pmatrix}}
\newcommand{\bmm}{\begin{matrix}}
\newcommand{\emm}{\end{matrix}}

\begin{document}

\title{Ground State Degeneracy in the Levin-Wen Model
for Topological Phases}

\author{Yuting Hu}
\email{yuting@physics.utah.edu}
\affiliation{Department of Physics and Astronomy,
  University of Utah, Salt Lake City, UT 84112, USA}
\author{Spencer D. Stirling}
\email{stirling@physics.utah.edu}
\affiliation{Department of Physics and Astronomy,
  University of Utah, Salt Lake City, UT 84112, USA}
\affiliation{Department of Mathematics,
  University of Utah, Salt Lake City, UT 84112, USA}
\author{Yong-Shi Wu}
\email{wu@physics.utah.edu} \affiliation{Department of Physics and
Center for Field Theory and Particle Physics, Fudan University,
Shanghai 200433, China} \affiliation{Department of Physics and
Astronomy, University of Utah, Salt Lake City, UT 84112, USA}

\date{\today}

\begin{abstract}
We study properties of topological phases by calculating the
ground state degeneracy (GSD) of the $2$d Levin-Wen (LW) model.
Here it is explicitly shown that the GSD depends only on the
spatial topology of the system. Then we show that the ground
state on a sphere is always non-degenerate. Moreover, we study
an example associated with a quantum group, and show that the
GSD on a torus agrees with that of the doubled Chern-Simons
theory, consistent with the conjectured equivalence between
the LW model associated with a quantum group and the doubled
Chern-Simons theory.

\end{abstract}


\maketitle

\section{Introduction}

In recent years two-dimensional topological phases
have received growing attention from the science community.
They represent a novel class of quantum matter
at zero temperature whose bulk properties are robust against weak
interactions and disorders.  Topological phases may be divided
into two families: \textit{doubled} (with time-reversal symmetry,
or TRS, preserved), and \textit{chiral} ( with TRS broken). Either
type may be exploited to do fault-tolerant (or topological)
quantum computing \cite{Kitaev,FKLW,NSSFD,Wang}.

\textit{Chiral} topological phases were first discovered in integer
and fractional quantum Hall (IQH and FQH) liquids. Mathematically,
their effective low-energy description is given by Chern-Simons
theory \cite{ZhHK} or (more generally) topological quantum field
theory (TQFT) \cite{Witten89}. One characteristic property of FQH
states is ground state degeneracy (GSD), which depends only on
the spatial topology of the system \cite{TW84,NTW85,WN90} and is
closely related to fractionization \cite{HKW90,HKW91,SKW06} of
quasiparticle quantum numbers, including fractional (braiding)
statistics \cite{Wilczek82,Wu84}. In some cases the GSD has been
computed in refs. \cite{WZ91,SKW06}.

Chern-Simons theories are formulated in the continuum and have no
lattice counterpart. Doubled topological phases, on the other
hand, do admit a discrete description. The first known example was
Kitaev's toric code model \cite{Kitaev}.

More recently, Levin and Wen (LW) \cite{LW} constructed a discrete
model to describe a large class of doubled phases. Their
original motivation was to generate ground states that exhibit the
phenomenon of string-net condensation \cite{Wen03} as a physical
mechanism for topological phases. The LW model is defined on a
trivalent lattice (or graph) with an exactly soluble Hamiltonian.
The ground states in this model can be viewed as the fixed-point
states of some renormalization group flow \cite{CGW,WEN11}.  These
fixed-point states look the same at all length scales and have no
local degrees of freedom.

The LW model is believed to be a Hamiltonian version of the
Turaev-Viro topological quantum field theory (TQFT) in three
dimensional spacetime \cite{Turaev94,KMR,Wang} and, in
particular cases, discretized version of \textit{doubled} 
Chern-Simons theory \cite{FNSWW,Simon}. Like Kitaev's toric 
code model \cite{Kitaev}, we expect that the subspace of 
degenerate ground states in the LW model can be used as a 
fault-tolerant code for quantum computation.

In this paper we report the results of a recent study on the GSD
of the LW model formulated on a (discretized) closed oriented
surface $M$. Usually the GSD is examined as a topological
invariant\cite{Turaev94,KMR,Simon} of the 3-manifold $S^1\times M$.
In a Hamiltonian approach accessible to physicists, we will
explicitly demonstrate that the GSD in the LW model depends only
on the topology of $M$ on which the system lives and, therefore,
is a topological invariant of the surface $M$. We also show that
the ground state of any LW Hamiltonian on a sphere is always
non-degenerate. Moreover, we examine the LW model associated with
quantum group $SU_k(2)$, which is conjectured to be equivalent to
the doubled Chern-Simons theory with gauge group $SU(2)$ at level
$k$, and compute the GSD on a torus. Indeed we find an agreement
with that in the corresponding doubled Chern-Simons theory
\cite{Witten89,RT}. This supports the above-mentioned conjectured
equivalence between the doubled Chern-Simons theory and the LW
model, at least in this particular case.

The paper is organized as follows. In Section II we present
the basics of the LW model, easy to read for newcomers.
In Section III topological properties of the ground states
are studied, and the topological invariance of their
degeneracy is shown explicitly. In section IV we demonstrate
how to calculate the GSD in a general way. In section V we
provide examples for the calculation particularly on a torus.
Section VI is devoted to summary and discussions. The detailed
computation of the GSD is presented in the appendices.

\section{The Levin-Wen model}

Start with a fixed (connected and directed) trivalent graph
$\Gamma$ which discretizes a closed
oriented surface $M$ (such as a torus). To each edge in the graph
we assign a string type $j$, which runs over a finite set
$j=0,1,...,N$.  Each string type $j$ has a ``conjugate'' $j^*$
that describes the effect of reversing the edge direction.  For
example $j$ may be an irreducible representation of a finite group
or (more generally) a quantum group \cite{Kassel}.

Let us associate to each string type $j$ a quantum dimension
$d_j$, which is a positive number for the Hamiltonian we define
later to be hermitian. To each triple of strings $\{i,j,k\}$ we
associate a \textit{branching rule} $\delta_{ijk}$ that equals $1$
if the triple is ``allowed'' to meet at a vertex, $0$ if not (in
representation language the tensor product $i\otimes j\otimes k$
either contains the trivial representation or not). This data must
satisfy (here $D=\sum_{j}d^2_j$)
\begin{align}
\label{dimcond}
\sum_{k}d_{k}\delta_{ijk^{*}}=d_{i}d_{j}\nonumber\\
\sum_{ij}d_{i}d_{j}\delta_{ijk^{*}}=d_{k}D
\end{align}
$j=0$ is the unique ``trivial'' string type,  satisfying $0^*=0$
and $\delta_{0jj^*}=1, \delta_{0ji^*}=0$ if $i\neq{j}$.

The Hilbert space is spanned by all configurations of all possible
string types $j$ on edges. The Hamiltonian is a sum of some
mutually-commuting projectors
$H:=-\sum_{v}\hat{Q}_v-\sum_p\hat{B}_p$ (one for each vertex $v$
and each plaquette $p$). Here each projector
$\hat{Q}_v=\delta_{ijk}$ with $i,j,k$ on the edges incoming to the
vertex $v$. $\hat{Q}_v=1$ enforces the branching rule on $v$.
Throughout the paper we work on the subspace of states in which
$\hat{Q}_v=1$ for all vertices. Each projector $\hat{B}_p$ is a
sum $D^{-1}\sum_{s}d_{s}\hat{B}^s_p$ of operators that have matrix
elements (on a hexagonal plaquette for example)
\begin{align}
\label{Bps}
&\Biggl\langle\bmm\includegraphics[height=0.6in]{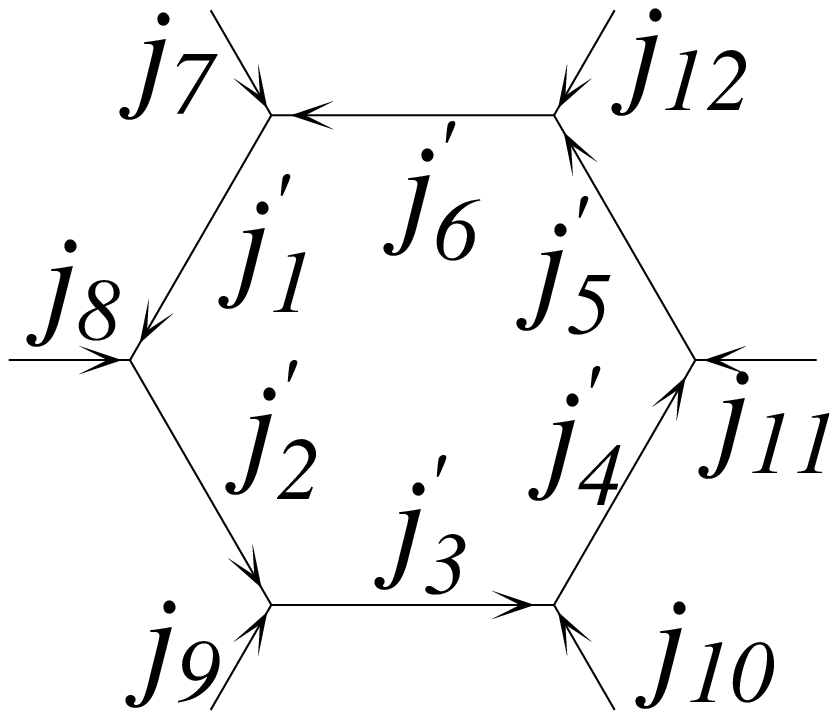}\emm\Biggr|
\hat{B}_p^s
\Biggl|\bmm\includegraphics[height=0.6in]{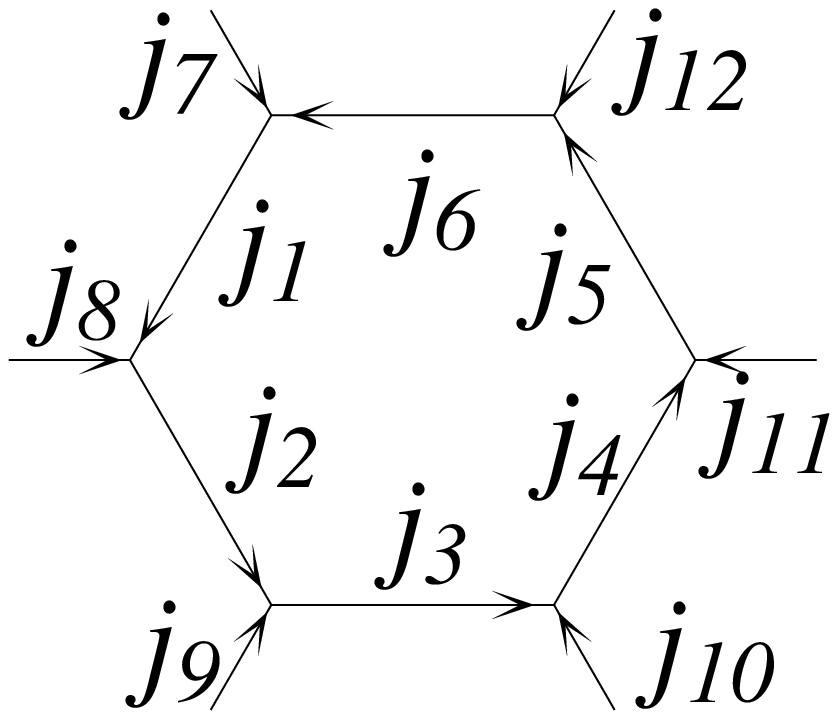}\emm\Biggr\rangle\nonumber\\
&=
v_{j_1}v_{j_2}v_{j_3}v_{j_4}v_{j_5}v_{j_6}v_{j'_1}v_{j'_2}v_{j'_3}v_{j'_4}v_{j'_5}v_{j'_6}\\
&G^{{j_7}{j^*_1}{j_6}}_{{s^*}{j'_6}{j'^*_1}}G^{{j_8}{j^*_2}{j_1}}_{{s^*}{j'_1}{j'^*_2}}
G^{{j_9}{j^*_3}{j_2}}_{{s^*}{j'_2}{j'^*_3}}G^{{j_{10}}{j^*_4}{j_3}}_{{s^*}{j'_3}{j'^*_4}}
G^{{j_{11}}{j^*_5}{j_4}}_{{s^*}{j'_4}{j'^*_5}}G^{{j_{12}}{j^*_6}{j_5}}_{{s^*}{j'_5}{j'^*_6}}
\nonumber
\end{align}
Here $v_j=\sqrt{d_j}$ is real.  The symmetrized $6j$ symbols\cite{WEN11} $G$
are complex numbers that satisfy
\begin{align}
\label{6jcond}
&\text{symmetry:}&G^{ijm}_{kln}=G^{mij}_{nk^{*}l^{*}}
=G^{klm^{*}}_{ijn^{*}}=(G^{j^*i^*m^*}_{l^*k^*n})^*\nonumber\\
&\text{pentagon id:}
&\sum_{n}{d_{n}}G^{mlq}_{kp^{*}n}G^{jip}_{mns^{*}}G^{js^{*}n}_{lkr^{*}}
=G^{jip}_{q^{*}kr^{*}}G^{riq^{*}}_{mls^{*}}\nonumber\\
&\text{orthogonality:}
&\sum_{n}{d_{n}}G^{mlq}_{kp^{*}n}G^{l^{*}m^{*}i^{*}}_{pk^{*}n}
=\frac{\delta_{iq}}{d_{i}}\delta_{mlq}\delta_{k^{*}ip}
\end{align}

For example, these conditions are known to be satisfied \cite{LW}
if we take the string types $j$ to be all irreducible
representations of a finite group, $d_j$ to be the dimension of
corresponding representation space, and $G$ to be the symmetrized
Racah $6j$ symbols for the group. In this case the LW model can be
mapped \cite{BA} to Kitaev's quantum double model \cite{Kitaev}.
More general sets of data $\{G,d,\delta\}$ can be derived from
quantum groups (or Hopf algebras) \cite{Kassel}. We will discuss
such a case later using the quantum group $SU_k(2)$ ($k$ being the
level).

\section{Ground states}

Any ground state $|\Phi\rangle$
(there may be many) must be a simultaneous $+1$ eigenvector for
all projectors $\hat{Q}_v$ and $\hat{B}_p$. In this section we
demonstrate the topological properties of the ground states on a
closed surface with non-trivial topology.

Let us begin with \textit{any two} arbitrary trivalent graphs
$\Gamma^{(1)}$ and $\Gamma^{(2)}$ discretizing the same surface
(e.g., a torus). If we compare the LW models based on these two
graphs, respectively, then immediately we see that the Hilbert
spaces are quite different from each other (they have different
sizes in general).

\begin{figure}[t!]
\begin{center}
  \quad\includegraphics[height=0.75in]{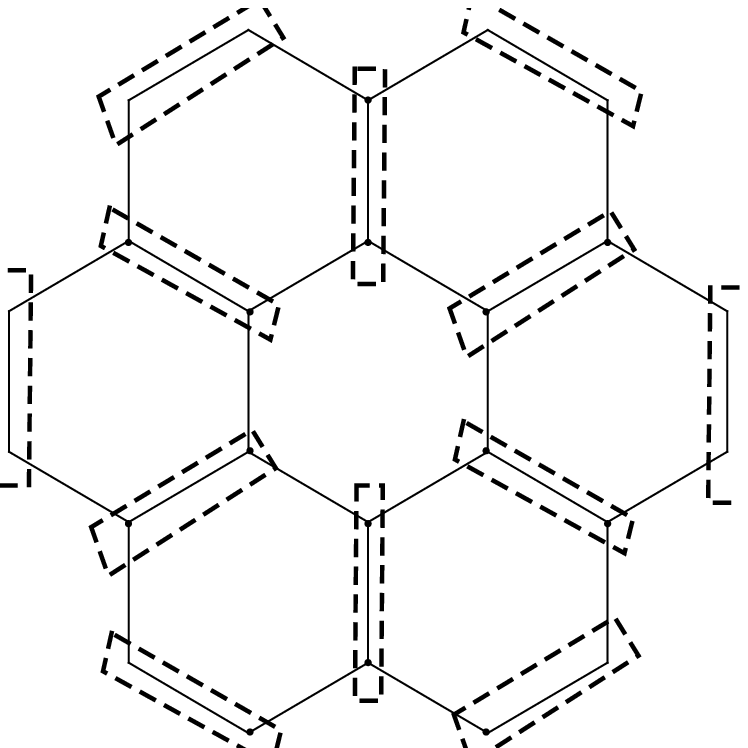}\qquad
  \quad\includegraphics[height=0.75in]{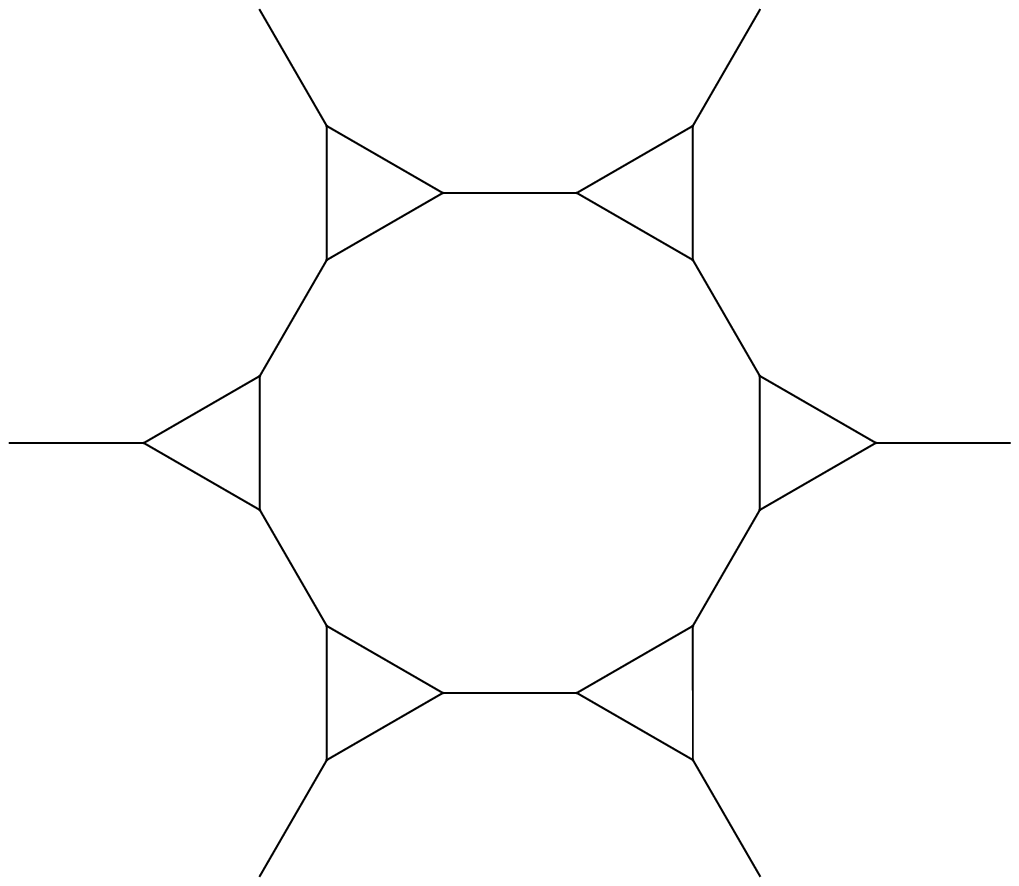}\qquad\quad
  \includegraphics[height=0.75in]{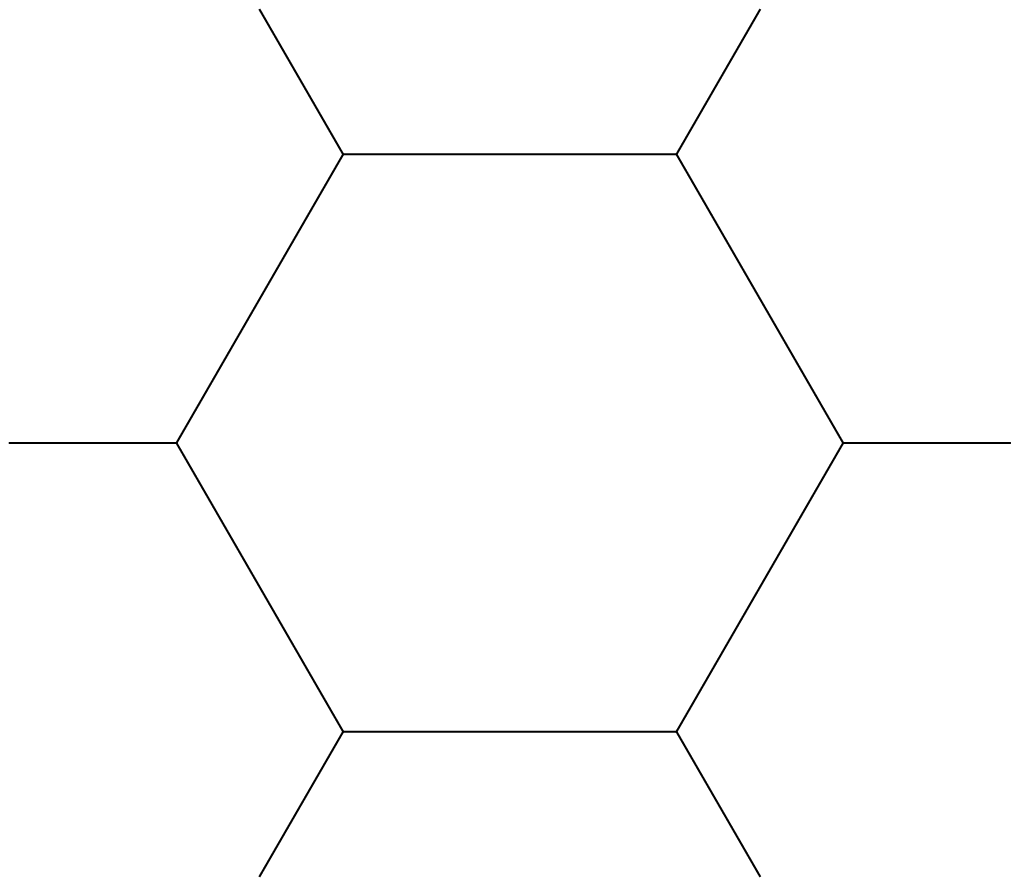}
  \newline
  $\Gamma^{(1)}\qquad\qquad\qquad\qquad\Rightarrow\qquad\qquad\qquad
  \qquad\Gamma^{(2)}$
  \caption{Given any two trivalent graphs $\Gamma^{(1)}$ and
  $\Gamma^{(2)}$ discretizing the same surface, we can always mutate
  $\Gamma^{(1)}$ to $\Gamma^{(2)}$ by a composition of elementary $f$
  moves. In general $\Gamma^{(1)}$ and $\Gamma^{(2)}$ are not required
  to be regular lattices.  These diagrams happen to be the
  same as \cite{Gu09}, but in a slightly different context.}
  \label{fig:reducedgraphsidea}
\end{center}
\end{figure}

However, we may mutate between any two given trivalent graphs
$\Gamma^{(1)}$ and $\Gamma^{(2)}$ by a composition of the
following elementary moves \cite{Pachner} (see also Fig
\ref{fig:reducedgraphsidea}
):
\begin{align}
&f_1. \bmm\includegraphics[height=0.4in]{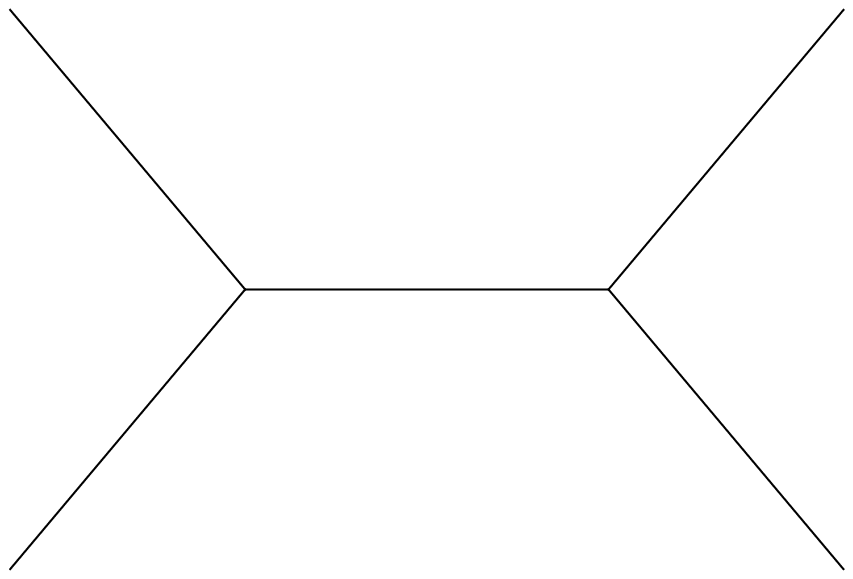}\emm\Rightarrow
\bmm\includegraphics[height=0.4in]{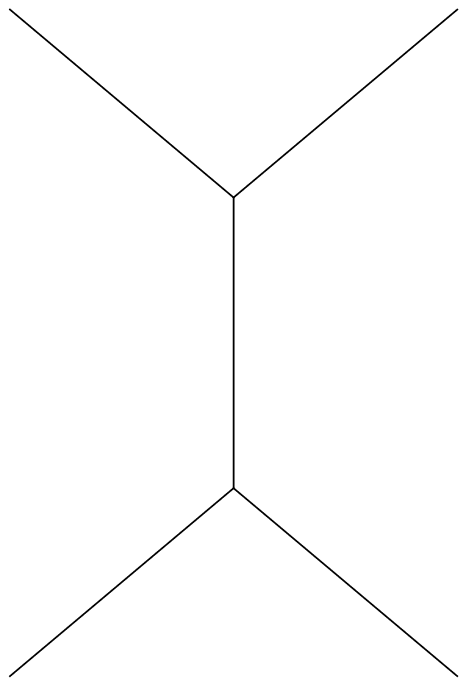}\emm\text{, for any edge;}
\nonumber\\
&f_2. \bmm\includegraphics[height=0.4in]{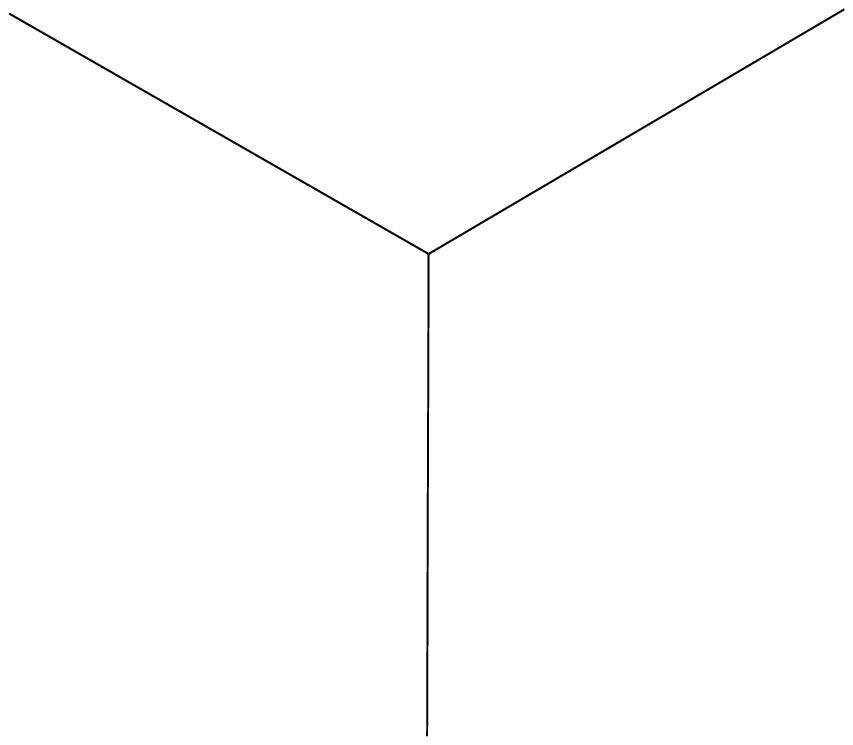}\emm\Rightarrow
\bmm\includegraphics[height=0.4in]{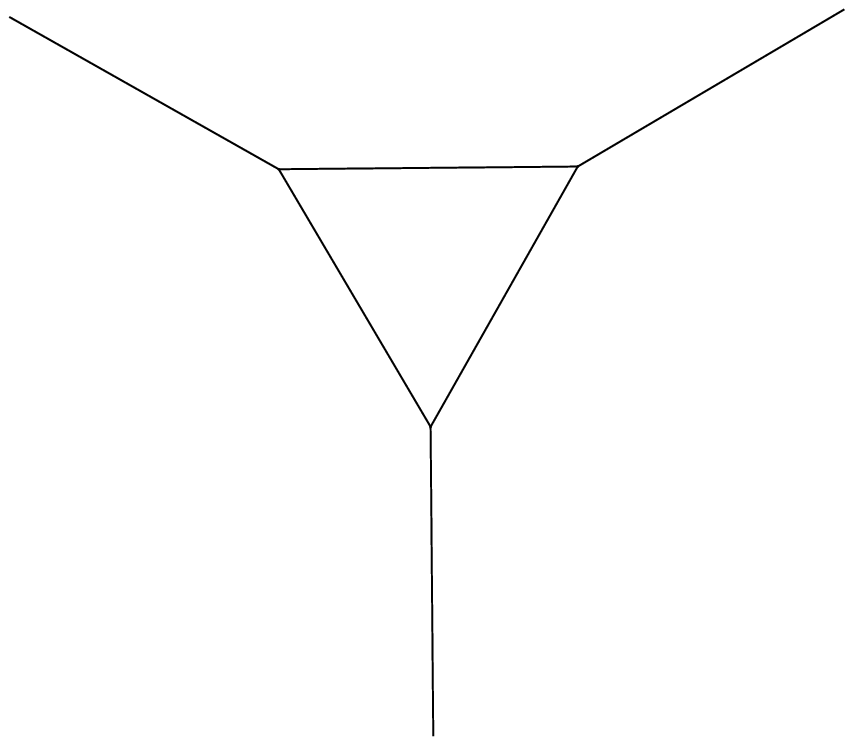}\emm\text{, for any vertex.}
\nonumber\\
&f_3. \bmm\includegraphics[height=0.4in]{def2b.eps}\emm\Rightarrow
\bmm\includegraphics[height=0.4in]{def2a.eps}
\emm\text{, for any triangle structure.}
\nonumber
\end{align}


Suppose we are given a sequence of elementary $f$ moves that
connects two graphs $\Gamma^{(1)}\rightarrow\Gamma^{(2)}$. We now
construct a linear transformation
$\mathcal{H}^{(1)}\rightarrow\mathcal{H}^{(2)}$ between the two
Hilbert spaces. This is defined by associating linear maps to each
elementary $f$ move:
\begin{align}
\label{T1T2T3}
&\hat{T}_1:\Biggl|\bmm\includegraphics[height=0.4in]{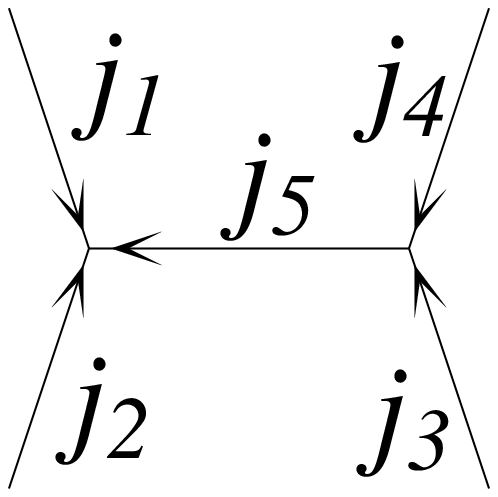}\emm\Biggr\rangle\rightarrow\sum_{j'_{5}}v_{j_5}v_{j'_5}G^{j_{1}j_{2}j_{5}}_{j_{3}j_{4}j'_{5}}
\Biggl|\bmm\includegraphics[height=0.4in]{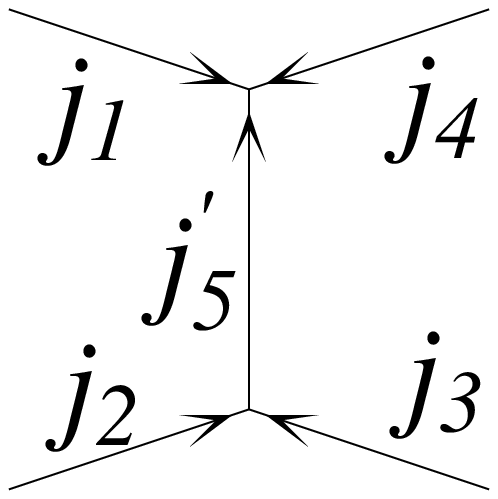}\emm\Biggr\rangle\nonumber\\
&\hat{T}_2:\Biggl|\bmm\includegraphics[height=0.4in]{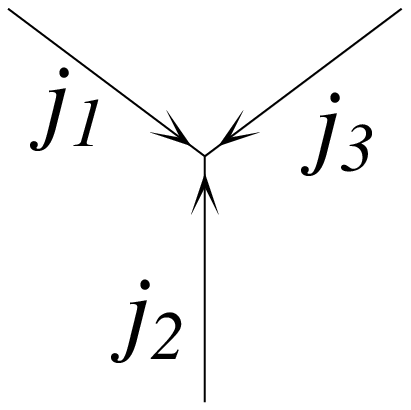}\emm\Biggr\rangle\rightarrow\sum_{j_{4}j_{5}j_{6}}\frac{v_{j_4}v_{j_5}v_{j_6}}
{\sqrt{D}}G^{j_{2}j_{3}j_{1}}_{j^*_{6}j_{4}j^*_{5}}
\Biggl|\bmm\includegraphics[height=0.4in]{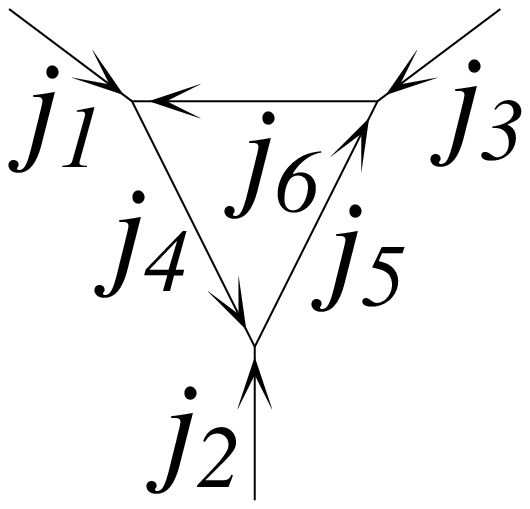}\emm\Biggr\rangle\nonumber\\
&\hat{T}_3:\Biggl|\bmm\includegraphics[height=0.4in]{Y2.eps}
\emm\Biggr\rangle\rightarrow
\frac{v_{j_4}v_{j_5}v_{j_6}}{\sqrt{D}}
G^{j^*_{3}j^*_{2}j^*_{1}}_{j^*_{4}j_{6}j^*_{5}}
\Biggl|\bmm\includegraphics[height=0.4in]{Y1.eps}\emm\Biggr\rangle
\end{align}

The mutation transformations between $\mathcal{H}^{(1)}$ and $\mathcal{H}^{(2)}$
are constructed by a composition of
these elementary maps. As a special example, the operator
$\hat{B}_p=D^{-1}\sum_{s}d_s\hat{B}_p^s$ is such a transformation.
In fact, on the particular triangle plaquette $p$ as in \eqref{T1T2T3},
we have $\hat{B}_{p=\triangledown}=\hat{T}_2\hat{T}_3$, by using
the pentagon identity in \eqref{6jcond}.


Mutation transformations are unitary on the ground states.
To see this, we only need to check that the elementary maps
$\hat{T}_1$, $\hat{T}_2$, and $\hat{T}_3$ are unitary.
First note that the following relations hold:
$\hat{T}_1^\dagger=\hat{T}_1$, $\hat{T}_2^\dagger=\hat{T}_3$,
and $\hat{T}_3^\dagger=\hat{T}_2$.  We emphasize that these are maps between the
Hilbert spaces on two different graphs.
For example, we check $\hat{T}_1^\dagger=\hat{T}_1$ by comparing matrix elements
\begin{align}
  \Biggl\langle\bmm\includegraphics[height=0.4in]{X1.eps}\emm\Biggr\vert
  \hat{T}_1^\dagger
  \Biggl\vert\bmm\includegraphics[height=0.4in]{X2.eps}\emm\Biggr\rangle
  \equiv
  &\Biggl(\Biggl\langle\bmm\includegraphics[height=0.4in]{X2.eps}\emm\Biggr\vert
  \hat{T}_1
  \Biggl\vert\bmm\includegraphics[height=0.4in]{X1.eps}\emm\Biggr\rangle\Biggr)^*
  \nonumber\\
  =&v_{j_5}v_{j'_5}\left(G^{j_1 j_2 j_5}_{j_3 j_4 j'_5}\right)^*
  \nonumber\\
  =&v_{j'_5}v_{j_5}G^{j_4 j_1 j'_5}_{j_2 j_3 j^*_5}
  \nonumber\\
  =&\Biggl\langle\bmm\includegraphics[height=0.4in]{X1.eps}\emm\Biggr\vert
  \hat{T}_1
  \Biggl\vert\bmm\includegraphics[height=0.4in]{X2.eps}\emm\Biggr\rangle
\end{align}
where in the third equality we used the symmetry condition in \eqref{6jcond}.

Similarly, for $\hat{T}_2^\dagger=\hat{T}_3$ (or $\hat{T}_3^\dagger=\hat{T}_2$), we have
\begin{align}
  \Biggl\langle\bmm\includegraphics[height=0.4in]{Y1.eps}\emm\Biggr\vert
  \hat{T}_2^\dagger
  \Biggl\vert\bmm\includegraphics[height=0.4in]{Y2.eps}\emm\Biggr\rangle
  \equiv
  &\Biggl(\Biggl\langle\bmm\includegraphics[height=0.4in]{Y2.eps}\emm\Biggr\vert
  \hat{T}_2
  \Biggl\vert\bmm\includegraphics[height=0.4in]{Y1.eps}\emm\Biggr\rangle\Biggr)^*
  \nonumber\\
  =&\frac{v_{j_4}v_{j_5}v_{j_6}}
   {\sqrt{D}}\left(G^{j_{2}j_{3}j_{1}}_{j^*_{6}j_{4}j^*_{5}}\right)^*
  \nonumber\\
  =&\frac{v_{j_4}v_{j_5}v_{j_6}}{\sqrt{D}}G^{j^*_{3}j^*_{2}j^*_{1}}_{j^*_{4}j_{6}j^*_{5}}
  \nonumber\\
  =&\Biggl\langle\bmm\includegraphics[height=0.4in]{Y1.eps}\emm\Biggr\vert
  \hat{T}_3
  \Biggl\vert\bmm\includegraphics[height=0.4in]{Y2.eps}\emm\Biggr\rangle
\end{align}

Now we verify unitary.
First, $\hat{T}_1^\dagger\hat{T}_1=\text{id}$ and
$\hat{T}_2^\dagger\hat{T}_2=\hat{T}_3\hat{T}_2=\text{id}$
by the orthogonality condition in \eqref{6jcond}
(note that, since we have not used any information about the ground states in this argument,
$\hat{T}_1$ and $\hat{T}_2$ are unitary on the entire Hilbert space).
For unitary of $\hat{T}_3$ we check $\hat{T}_3^\dagger\hat{T}_3=\hat{T}_2\hat{T}_3=1$.
The last equality only holds on the ground states since we have already seen
that $\hat{T}_2\hat{T}_3=\hat{B}_{p=\triangledown}$ and
$\hat{B}_{p=\triangledown}=1$ only on the ground states.

As another consequence of the above relations,
the Hamiltonian is hermitian since all $\hat{B}_p$'s consist
of elementary $\hat{T}_1$, $\hat{T}_2$, and $\hat{T}_3$ maps. Particularly,
on a triangle plaquette, we have
$\hat{B}_{p=\triangledown}^\dagger=(\hat{T}_2\hat{T}_3)^\dagger=
\hat{T}_3^\dagger\hat{T}_2^\dagger=\hat{T}_2\hat{T}_3=\hat{B}_{p=\triangledown}$.



The mutation transformations serve as the symmetry transformations
in the ground states. If $\vert\Phi\rangle$ is a ground state then
$\hat{T}\vert\Phi\rangle$ is also a ground state, where $\hat{T}$
is a composition of $\hat{T}_i$'s associated with elementary $f$
moves from $\Gamma^{(1)}$ to $\Gamma^{(2)}$. This is equivalent to
the condition $\hat{T}(\prod_p \hat{B}_p)=(\prod_{p'}
\hat{B}_{p'})\hat{T}$, which can be verified by the conditions
in \eqref{6jcond}. (Here $p$ and $p'$ run over the plaquettes
on $\Gamma^{(1)}$ and $\Gamma^{(2)}$, respectively.
Also note that the $\hat{B}_p$'s are mutually-commuting projectors, i.e.,
$\hat{B}_p\hat{B}_p=\hat{B}_p$, and thus
$\prod_p\hat{B}_p$ is the projector that
projects onto the ground states.)

These symmetry transformations look a little different from the
usual ones since they may transform between the Hilbert spaces
$\mathcal{H}^{(1)}$ and $\mathcal{H}^{(2)}$ on two different
graphs $\Gamma^{(1)}$ and $\Gamma^{(2)}$. In general,
$\Gamma^{(1)}$ and $\Gamma^{(2)}$ do not have the same number of
vertices and edges. And thus $\mathcal{H}^{(1)}$ and
$\mathcal{H}^{(2)}$ have different sizes. However, if we restrict
to the ground-state subspaces $\mathcal{H}^{(1)}_{0}$ and
$\mathcal{H}^{(2)}_{0}$, mutation transformations are invertible.
In fact, they are unitary as we have just shown.

The tensor equations on the $6j$ symbols in \eqref{6jcond}
give rise to a simple result: each mutation that preserves the
spatial topology of the two graphs
induces a unitary symmetry transformation.
During the mutations, local structures of the graphs
are destroyed, while the spatial topology of the
graphs is not changed. Correspondingly,
the local information of the ground states may be lost,
while the topological feature of the ground states is preserved.
In fact, any topological feature can be specified by a topological observable
$\hat{O}$ that is invariant under all mutation
transformations $\hat{T}$ from $\mathcal{H}^{(1)}$ to
$\mathcal{H}^{(2)}$: $\hat{O}'\hat{T}=\hat{T}\hat{O}$ (where
$\hat{O}$ is defined on the graph $\Gamma^{(1)}$ and $\hat{O}'$ on
$\Gamma^{(2)}$).

The symmetry transformations provides a way to characterize the
topological phase by a topological observable. In the next section we
will investigate the GSD as such an observable.


Let us end this section by remarking on uniqueness of the mutation transformations. There may be
many ways to mutate $\Gamma^{(1)}$ to $\Gamma^{(2)}$
using $f_1$, $f_2$ and $f_3$ moves. Each way determines a
corresponding transformation between the Hilbert spaces of ground
states, $\mathcal{H}^{(1)}_{0}$ and $\mathcal{H}^{(2)}_{0}$. It
turns out that all these transformations are actually the same if
the initial and final graphs $\Gamma^{(1)}$ to $\Gamma^{(2)}$ are
fixed, i.e., independent of which way we choose to mutate the
graph $\Gamma^{(1)}$ to $\Gamma^{(2)}$. This means that the ground
state Hilbert spaces on different graphs can be identified (up to
a mutation transformation) and all graphs are equally good.

One consequence of the uniqueness of the mutation tranformation is
that the degrees of freedom in the ground states do not depend on
the specific structure of the graph. In this sense, the LW model
is the Hamiltonian version of some discrete TQFT (actually,
Turaev-Viro type TQFT, see\cite{KMR}). The fact that the degrees
of freedom of the ground states depend only on the topology of the
closed surface $M$ is a typical characteristic of topological
phases \cite{TW84,NTW85,WN90,WZ91,SKW06}.

\section{ Ground state degeneracy}

In this section we investigate the simplest nontrivial topological observable, the GSD.
Since $\prod_p\hat{B}_p$ is the projector that projects onto the ground states,
taking a trace computes
$\text{GSD}=\text{tr}(\prod_{p}\hat{B}_p)$.

We can show that GSD is a topological invariant.
Namely, in the previous section we mentioned that, by using \eqref{6jcond},
$\prod_{p}\hat{B}_p$ is invariant under any mutation $\hat{T}$ between
the Hilbert spaces $\mathcal{H}^{(1)}$ and $\mathcal{H}^{(2)}$ :
$\hat{T}^\dagger(\prod_{p'}\hat{B}_{p'})\hat{T}=\prod_{p}\hat{B}_{p}$.
Taking a trace of both sides leads to
$\text{tr}^\prime(\prod_{p'}\hat{B}_{p'})=\text{tr}(\prod_{p}\hat{B}_p)$, where the
traces are evaluated on $\mathcal{H}^{(2)}$ and $\mathcal{H}^{(1)}$ respectively.

\begin{figure}[t!]
\begin{center}
  (a)\includegraphics[height=0.8in]{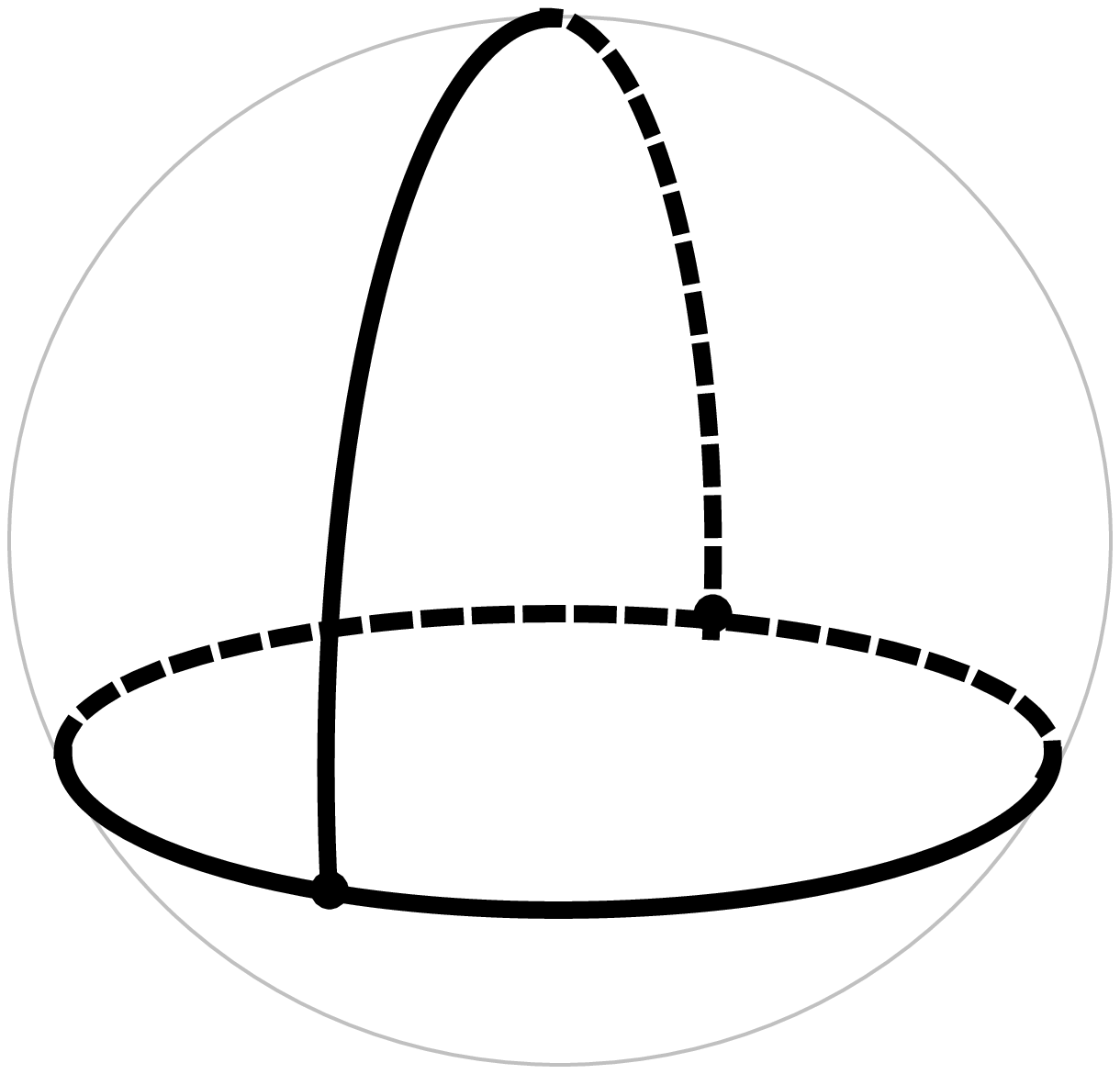}\quad\quad{(b)}\includegraphics[height=0.8in]{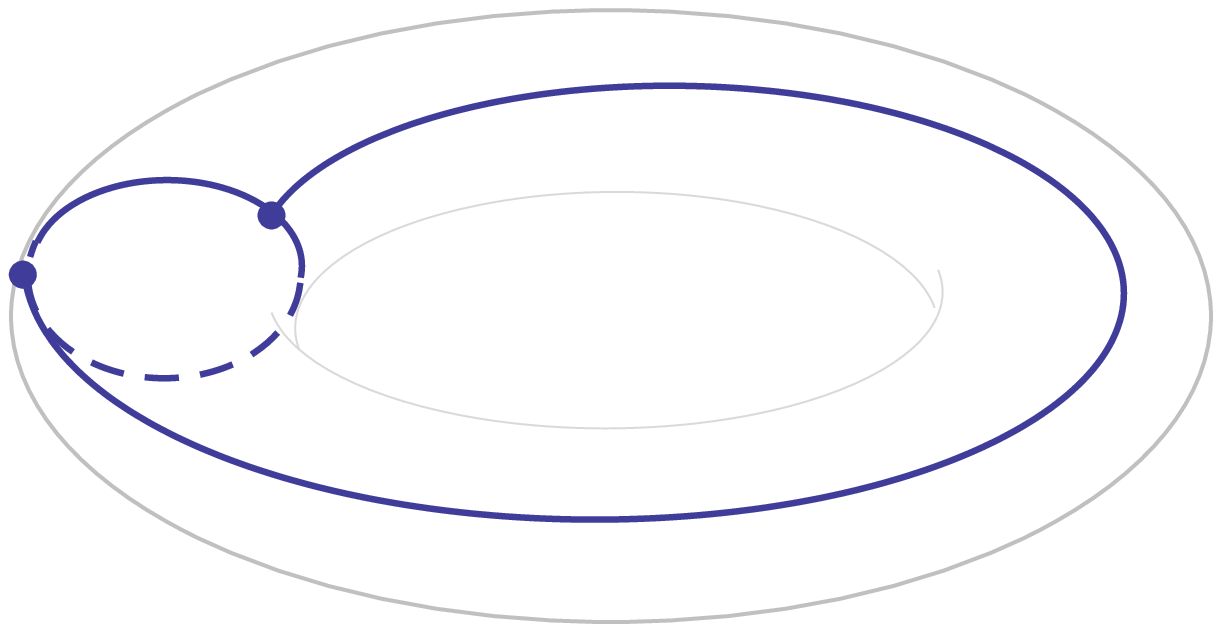}
  \caption{All trivalent graphs can be reduced to their simplest structures by compositions of elementary $f$ moves.
    (a) on a sphere: 2 vertices, 3 edges,
  and 3 plaquettes. (b) on a torus: 2 vertices, 3 edges, and 1 plaquette.}
  \label{fig:reducedgraphs}
\end{center}
\end{figure}

The independence of the GSD on the local structure of the graphs provides a practical algorithm
for computing the $\text{GSD}$, since we may always use the
simplest graph (see Fig \ref{fig:reducedgraphs} and examples in
the next section).

Expanding the GSD explicitly in terms of
$6j$ symbols using \eqref{Bps} we obtain
\begin{align}
\label{GSD}
&\text{GSD}=\sum_{{j_1}{j_2}{j_3}{j_4}{j_5}{j_6}...}\Biggl\langle\bmm\includegraphics[height=0.4in]{X1.eps}\emm
\Biggr|(\prod_{p}\hat{B}_p)\Biggl|\bmm\includegraphics[height=0.4in]{X1.eps}\emm\Biggr\rangle\nonumber\nonumber\\
&=D^{-P}\sum_{{s_1}{s_2}{s_3}{s_4}...}d_{s_1}d_{s_2}d_{s_3}d_{s_4}...\nonumber\\
&\quad\sum_{{j^\prime_1}{j^\prime_2}{j^\prime_3}{j^\prime_4}{j^\prime_5}...}
d_{j^\prime_1}d_{j^\prime_2}d_{j^\prime_3}d_{j^\prime_4}d_{j^\prime_5}...
\sum_{{j_1}{j_2}{j_3}{j_4}{j_5}...}d_{j_1}d_{j_2}d_{j_3}d_{j_4}d_{j_5}...\nonumber\\
&\quad
\left({G}^{{j_2}{j_5}{j_1}}_{{s^*_1}{j^\prime_1}{j^\prime_5}}
G^{{j^\prime_1}{j_2}{j^\prime_5}}_{{s^*_2}{j_5}{j^\prime_2}}
G^{{j_5}{j^\prime_1}{j^\prime_2}}_{{s^*_3}{j_2}{j_1}}\right)
\left(G^{{j_3}{j_4}{j^*_5}}_{{s^*_1}{j^{\prime*}_5}{j^\prime_4}}
G^{{j^\prime_4}{j^{\prime*}_5}{j_3}}_{{s^*_2}{j^\prime_3}{j^*_5}}
G^{{j^*_5}{j^\prime_3}{j^\prime_4}}_{{s^*_4}{j_4}{j_3}}\right)...
\end{align}
The formula needs some explanation.  $P$ is the total number of
plaquettes of the graph. Each plaquette $p$ contributes a
summation over $s_p$ together with a factor of
$\frac{d_{s_p}}{D}$. In the picture in \eqref{GSD} the top
plaquette is being operated on first by $\hat{B}^{s_1}_{p_1}$,
next the bottom plaquette by $\hat{B}^{s_2}_{p_2}$, third the left
plaquette by $\hat{B}^{s_3}_{p_3}$, and finally the right
plaquette by $\hat{B}^{s_4}_{p_4}$.  Although ordering of the
$\hat{B}^{s}_{p}$ operators is not important (since all
$\hat{B}_p$'s commute with each other), it is important to make an
ordering choice (for all plaquettes on the graph) \textit{once and
for all}.

Each edge $e$ contributes a summation over $j_e$ and $j^\prime_e$
together with a factor of $d_{j_e}d_{j^\prime_e}$. Each vertex
contributes three $6j$ symbols.

The indices on the $6j$ symbols work as follows: since each vertex
borders three plaquettes where $\hat{B}^{s}_{p}$'s are being
applied, we pick up a $6j$ symbol for each corner.  However,
ordering is important: because we have an overall ordering of
$\hat{B}^{s}_{p}$'s, at each vertex we get an induced ordering for
the $6j$ symbols. Starting with the $6j$ symbol furthest left we
have no primes on the top row.  The bottom two indices pick up
primes.  All of these variables (primed or not) are fed into the
next $6j$ symbol and the same rule applies: the bottom two indices
pick up a prime with the convention $()^{\prime\prime}=()$.


By the calculation of the GSD, we have characterized a topological
property of the phase using local quantities living on a graph
discretizing $M$ of nontrivial topology.

\section{Examples}

\noindent{\sl(1) On a sphere.} To
calculate the GSD, we need to input the data
$\{G^{ijm}_{kln},d_j,\delta_{ijm}\}$ and evaluate the trace in
\eqref{GSD}. We start by computing the GSD in the simplest case of
a sphere.

Let's consider the simplest graph as in Fig.
\ref{fig:reducedgraphs}(a). We show in Appendix A that the
ground state is non-degenerate on
the sphere without referring to any specific structure in the
model: $\text{GSD}^\text{sphere}=1$. In fact, for more general
graphs one can write down \cite{Gu09} the ground state as
$\prod_{p}\hat{B}_{p}|0\rangle$ up to a normalization factor,
where in $|0{\rangle}$ all edges are labeled by string type 0.

We notice that the GSD on the open disk (which is topologically
the same as the $2$d plane) can be studied using the same
technique. This is because the open disk can be obtained by
puncturing the sphere in Fig \ref{fig:reducedgraphs}(a) at the
bottom. Although this destroys the bottom plaquette, we notice
that the constraint $\hat{B}_p=1$ from the bottom plaquette is
automatically satisfied as a consequence of the same constraint on
all other plaquettes. The fact that
$\text{GSD}^{\text{sphere}}(=\text{GSD}^{\text{disk}})=1$
indicates the non-chiral topological order in the LW model.

\noindent{\sl(2) Quantum double model.} When the data are
determined by representations of a finite group $G$, the LW model
is mapped to Kitaev's quantum double model\cite{Kitaev,BA}. The
ground states corresponds one-to-one to the flat
$G$-connections\cite{Kitaev}. The GSD is
\begin{equation}
\label{GSDQD}
\text{GSD}_{\text{QD}}=\left|\frac{\text{Hom}(\pi_{1}(M),G)}{G}\right|
\end{equation}
where $\text{Hom}(\pi_{1}(\mathcal{M}),G)$ is the space of
homomorphisms from the fundamental group $\pi_1(M)$ to $G$, and
$G$ in the quotient acts on this space by conjugation.

In particular, the GSD \eqref{GSDQD} on a torus is
\begin{equation}
  \label{torusGSDQD}
  \text{GSD}^{\text{torus}}_{\text{QD}}
  =\left|\{(a,b)|a,b\in{G};aba^{-1}b^{-1}=e\}/\sim\right|
\end{equation}
where $\sim$ in the quotient is the equivalence by conjugation,
\begin{equation}
  (a,b)\sim (hah^{-1},hbh^{-1}) \quad \text{for all } h\in{G}
  \nonumber
\end{equation}

The number \eqref{torusGSDQD} is also the total number of
irreducible representations\cite{DPR} of the quantum double $D(G)$ of the
group $G$. On the other hand, the quasiparticles in the
model are classified\cite{Kitaev} by the quantum double $D(G)$.
Thus the GSD on a torus is equal to the number of particle species
in this example.

\noindent{\sl(3) $SU_k(2)$ structure on a torus.}  More
generally, on a torus any trivalent graph
can be reduced to the simplest one with two
vertices and three edges, as in Fig \ref{fig:reducedgraphs}(b). On
this graph the GSD consists of six local $6j$ symbols.
\begin{align}
\label{torusGSD}
&\text{GSD}=D^{-1}\sum_{sj_{1}j_{2}j_{3}j'_{1}j'_{2}j'_{3}}
d_{s}d_{j_{1}}d_{j_{2}}d_{j_{3}}d_{j'_{1}}d_{j'_{2}}d_{j'_{3}}\nonumber\\
&
\left(G^{j_{1}j_{2}j^{*}_{3}}_{{s}j'^{*}_{3}j'_{2}}
G^{{j'^*_3}{j_1}{j'_2}}_{{s}{j_2}{j'_1}}
G^{{j_2}{j'^*_3}{j'_1}}_{{s}{j_1}{j^*_3}}\right)
\left(G^{{j^*_2}{j_3}{j^*_1}}_{{s}{j'^*_1}{j'_3}}
G^{{j'_3}{j'^*_1}{j^*_2}}_{{s}{j'^*_2}{j^*_1}}
G^{{j^*_1}{j'^*_2}{j'_3}}_{{s}{j_3}{j^*_2}}
\right)
\end{align}

Now let us take the example using the quantum group $SU_k(2)$. It
is known that $SU_k(2)$ has $k+1$ irreducible representations, and
thus the GSD we calculate is finite. We take the string types to
be these representations, labeled as ${0,1,...,k}$, and the data
$\{G^{ijm}_{kln},d_j,\delta_{ijm}\}$ to be determined by these
representations (for more details, see\cite{RT,KR,MV}).

In Appendix B we show that in this case (for the LW model on a torus
with string types given by irreps of $SU_k(2)$)
we have $\text{GSD}=(k+1)^2$.  We argue this both analytically and numerically.

On the other hand, it is widely
believed that when the string types in the LW model are irreps
from a quantum group at level $k$, then the associated TQFT is given by
doubled Chern-Simons theory associated with the corresponding Lie
group at level ${\pm}k$\cite{Witten88,RT}. This equivalence tells
us that in this case the LW model can be viewed as a Hamiltonian
realization of the doubled Chern-Simons theory on a lattice, and
it provides an explicit picture of how the LW model describes
doubled topological phases.

Along these lines, our result is consistent
\cite{WZ} with the result $\text{GSD}_{CS}=k+1$ for Chern-Simons
$SU(2)$ theory at level $k$ on a torus.  This can be seen since the
Hilbert space associated to doubled Chern-Simons should be
the tensor product of two copies of Chern-Simons theory at level $\pm{k}$.

\section{Summary and Discussions}

In this paper, we studied the LW model that describes 2d
topological phases which do not break time-reversal symmetry.
By examining the 2d (trivalent) graphs with same topology
which are related to each other by a given finite set of
operations (Pachner moves), we developed techniques to deal
with topological properties of the ground states. Using them,
we have been able to show explicitly that the GSD is determined
only by the topology of the surface the system lives on,
which is a typical feature of topological phases. We also
demonstrated how to obtain the GSD from local data in a
general way. We explicitly showed that the ground state of
any LW Hamiltonian on a sphere is non-degenerate. Moreover,
the LW model associated with quantum group $SU_k(2)$ was
studied, and our result for the GSD on a torus is consistent
with the conjecture that the LW model associated with quantum
group is the realization of a doubled Chern-Simons theory on
a lattice or discrete graph.

Finally, let us indicate possible extension of the results to
more general cases. First, more generally in the LW model, an
extra discrete degree of freedom, labelled by an index
$\alpha$, may be put on the vertices. Then the branching rule
$\delta_{ijk}^{\alpha}$, when its value is $1$, may carry an extra
index $\alpha$. (In representation language this implies that
given irreducible representations $i$,$j$ and $k$, there may be
multiple inequivalent ways to obtain the trivial representation
from the tensor product of $i\otimes j\otimes k$. The index
$\alpha$ just labels these different ways.) The $6j$ symbols
accordingly carry more indices. (For more details see the first
Appendix in the original paper \cite{LW} of the LW model.) The
expression \eqref{GSD} for GSD is expected to be generalizable to
these cases. Secondly, the spatial manifold (e.g. a torus) on
which the graph is defined may carry non-trivial charge, e.g.
labelled by $i\bar{i}$ in the $SU_k(2)$ case. This corresponds to
having a so-called fluxon excitation (of type $i\bar{i}$) above
the original LW ground states. The lowest states of this subsector
in the LW model coincide with the ground states for the
Hamiltonian obtained by replacing the plaquette projector
$\hat{B}_p = D^{-1} \sum_j d_j \hat{B}_p^j$ with $\hat{B}_p =
D^{-1} \sum_j s_{ij} \hat{B}_p^j$, where $s_{ij}$ is the modular
$S$-matrix. (See Appendix B.)
The GSD in this case is computable too, but we leave this for a
future paper \cite{HSW2}.

\begin{acknowledgments}
YH thanks Department of Physics, Fudan University for warm
hospitality he received during a visit in summer 2010. YSW 
was supported in part by US NSF through grant No. PHY-0756958, 
No. PHY-1068558 and by FQXi.
\end{acknowledgments}

\appendix

\section{$\text{GSD}=1$ on a sphere}

In appendix, we derive $\text{GSD}=1$ on a sphere for a
general Levin-Wen model, without referring to any specific
structure of the data $\{d,\delta,G\}$. All we will use in the
derivation are the general properties in eq. \eqref{dimcond} and
eq. \eqref{6jcond}.

The simplest trivalent graph on a sphere has three plaquettes and
three edges, as illustrated in Fig. \ref{fig:reducedgraphs}(a).
Following the standard procedure as in \eqref{GSD}, the GSD is
expanded as
\begin{align}
\label{GSDsphere}
  &\text{GSD}^{\text{sphere}}
  =\sum_{j_1j_2j_3}
  \Biggl\langle\bmm
  {\scalefont{0.6}
  \begin{tikzpicture}[scale=0.6]
  \draw [<-,>=latex] (2,0) arc (0:180:1 and 0.8);
  \draw [->,>=latex] (0,0) arc (180:360:1 and 0.8);
  \draw [->,>=latex] (2,0) -- (0,0);
  \node at(1,1.) {$j_1$};
  \node at(1,-0.6) {$j_2$};
  \node at(1,0.2) {$j_3$};
  \end{tikzpicture}
  }
  \emm\Biggr\vert
  \hat{B}_{p_2}\hat{B}_{p_3}\hat{B}_{p_1}
    \Biggl\vert\bmm
  {\scalefont{0.6}
  \begin{tikzpicture}[scale=0.6]
  \draw [<-,>=latex] (2,0) arc (0:180:1 and 0.8);
  \draw [->,>=latex] (0,0) arc (180:360:1 and 0.8);
  \draw [->,>=latex] (2,0) -- (0,0);
  \node at(1,1.) {$j_1$};
  \node at(1,-0.6) {$j_2$};
  \node at(1,0.2) {$j_3$};
  \end{tikzpicture}
  }
  \emm\Biggr\rangle
  \nonumber\\
  =&\sum_{j_1j_2j_3}
  \Biggl\langle\bmm
  {\scalefont{0.6}
  \begin{tikzpicture}[scale=0.6]
  \draw [<-,>=latex] (2,0) arc (0:180:1 and 0.8);
  \draw [->,>=latex] (0,0) arc (180:360:1 and 0.8);
  \draw [->,>=latex] (2,0) -- (0,0);
  \node at(1,1.) {$j_1$};
  \node at(1,-0.6) {$j_2$};
  \node at(1,0.2) {$j_3$};
  \end{tikzpicture}
  }
  \emm\Biggr\vert
  \frac{1}{D}
  \sum_{t}d_{t}\hat{B}_{p_2}^t
  \nonumber\\
  &\quad\qquad
  \frac{1}{D}\sum_{s}d_{s}\hat{B}_{p_3}^s
  \frac{1}{D}\sum_{r}d_{r}\hat{B}_{p_1}^r
    \Biggl\vert\bmm
  {\scalefont{0.6}
  \begin{tikzpicture}[scale=0.6]
  \draw [<-,>=latex] (2,0) arc (0:180:1 and 0.8);
  \draw [->,>=latex] (0,0) arc (180:360:1 and 0.8);
  \draw [->,>=latex] (2,0) -- (0,0);
  \node at(1,1.) {$j_1$};
  \node at(1,-0.6) {$j_2$};
  \node at(1,0.2) {$j_3$};
  \end{tikzpicture}
  }
  \emm\Biggr\rangle
  \nonumber\\
  =&\sum_{j_1j_2j_3j'_1j'_2j'_3}
  \frac{1}{D}\sum_{r}d_r
  v_{j_1}v_{j_3}v_{j'_1}v_{j'_3}
  G_{r^*{j'_1}^*j'_3}^{j_2^*j_3j_1^*}G_{r^*{j'_3}^*{j'_1}}^{j_2j_1j_3^*}
  \nonumber\\
  &\qquad\qquad\frac{1}{D}\sum_{s}d_s
  v_{j'_1}v_{j_2}v_{j_1}v_{j'_2}
  G^{{j'_3}{j'_1}^*j_2^*}_{s^*{j'_2}^*j_1^*}G^{{j'_3}^*j_2{j'_1}}_{s^*j_1{j'_2}}
  \nonumber\\
  &\qquad\qquad\frac{1}{D}\sum_{t}d_t
  v_{j'_2}v_{j'_3}v_{j_2}v_{j_3}
  G^{j_1^*{j'_2}^*j'_3}_{t^*j_3j_2^*}G^{j_1{j'_3}^*{j'_2}}_{t^*j_2j_3^*}
\end{align}
where $\hat{B}_{p_1}$ is acting on the top bubble plaquette,
$\hat{B}_{p_2}$ on the bottom bubble plaquette, and
$\hat{B}_{p_3}$ on the rest plaquette outside the two bubbles.

All $6j$ symbols can be eliminated by using the orthogonality
condition in eq. \eqref{6jcond} three times, 
\begin{align}
  &\sum_{r}d_r
  G_{r^*{j'_1}^*j'_3}^{j_2^*j_3j_1^*}G_{r^*{j'_3}^*{j'_1}}^{j_2j_1j_3^*}
  =\frac{1}{d_{j_2}}\delta_{j'_1j_2{j'_3}^*}\delta_{j_1j_2j_3^*}
  \nonumber\\
  &\sum_{s}d_s
  G^{{j'_3}{j'_1}^*j_2^*}_{s^*{j'_2}^*j_1^*}G^{{j'_3}^*j_2{j'_1}}_{s^*j_1{j'_2}}
  =\frac{1}{d_{j'_3}}\delta_{j'_1j_2{j'_3}^*}\delta_{j_1j'_2{j'_3}^*}
  \nonumber\\
  &\sum_{t}d_t
  G^{j_1^*{j'_2}^*j'_3}_{t^*j_3j_2^*}G^{j_1{j'_3}^*{j'_2}}_{t^*j_2j_3}
  =\frac{1}{d_{j_1}}\delta_{j_1j_2j_3^*}\delta_{j_1j'_2{j'_3}^*}
\end{align}

and the GSD is a summation in terms of $\{d,\delta\}$:
\begin{align}
  \text{GSD}^{\text{sphere}}
  =\frac{1}{D^3}\sum_{j_1j_2j_3j'_1j'_2j'_3}d_{j'_1}d_{j'_2}d_{j_3}
  \delta_{j_1j_2j_3^*}\delta_{j'_1j_2{j'_3}^*}\delta_{j_1j'_2{j'_3}^*}
\end{align}

Summing over $j'_1$, $j'_2$, and $j_3$ using \eqref{dimcond}
finally leads to $\text{GSD}^{\text{sphere}}=1$.

\section{GSD on a torus for $SU_k(2)$}

Let us consider the example associated with  the quantum group
$SU_k(2)$ (with the level $k$ an positive integer) and calculate
the GSD on a torus.

There are $k+1$ string types, labeled as $j=0,1,2,...,k$. They are
the irreducible representations of $SU_k(2)$. The
quantum dimensions $d_j$ are required to be positive for all $j$,
in order that the Hamiltonian is hermitian. Explicitly, they are
\begin{align}
  \label{dD}
  & d_j=\frac{\sin{\frac{(j+1)\pi}{k+2}}}{\sin{\frac{\pi}{k+2}}}
  \nonumber\\
  & D=\sum_{j=0}^{k}{d_j^2}=\frac{k+2}{2\sin^2{\frac{\pi}{k+2}}}
\end{align}

The branching rule is $\delta_{rst}=1$ if
\begin{align}
\label{fusionrule} \Biggl\{
\begin{array}{l}
  r+s+t\text{ is even}  \\
  r+s\geq{t}, s+t\geq{r}, t+r\geq{s} \\
  r+s+t\leq{2k}
\end{array}
\Biggr.
\end{align}
and $\delta_{rst}=0$ otherwise. The explicit formula for the $6j$
symbol can be found in\cite{KR,MV}. However, we do not need the
detailed data of the $6j$ symbol in the following computation of
the GSD.

Let us start with formula in \eqref{torusGSD}, and reorder the
$6j$ symbols,
\begin{align}
\label{SU2GSDReOrganized}
\text{GSD}=&D^{-1}\sum_{sj_{1}j_{2}j_{3}j'_{1}j'_{2}j'_{3}} d_{s}
\left( v_{j_{1}}v_{j_3}v_{j'_1}v_{j'_3}
G^{{j^*_2}{j_3}{j^*_1}}_{{s^*}{j'^*_1}{j'_3}}
G^{{j_2}{j'^*_3}{j'_1}}_{{s^*}{j_1}{j^*_3}}\right)
\nonumber\\
&\quad\qquad\left( v_{j'_1}v_{j_2}v_{j_1}v_{j'_2}
G^{{j'_3}{j'^*_1}{j^*_2}}_{{s^*}{j'^*_2}{j^*_1}}
G^{{j'^*_3}{j_1}{j'_2}}_{{s^*}{j_2}{j'_1}}\right)
\nonumber\\
&\quad\qquad\left( v_{j'_2}v_{j'_3}v_{j_2}v_{j_3}
G^{{j^*_1}{j'^*_2}{j'_3}}_{{s^*}{j_3}{j^*_2}}
G^{j_{1}j_{2}j^{*}_{3}}_{{s^*}j'^{*}_{3}j'_{2}}\right)
\nonumber\\
=&D^{-1}\sum_{sj_{1}j_{2}j_{3}j'_{1}j'_{2}j'_{3}} d_{s} \left(
v_{j_{1}}v_{j_3}v_{j'_1}v_{j'_3}
G^{{j^*_2}{j_3}{j^*_1}}_{{s^*}{j'^*_1}{j'_3}}
G^{j_2^*j_1^*j_3}_{sj'_3{j'_1}^*}\right)
\nonumber\\
&\quad\qquad\left( v_{j'_1}v_{j_2}v_{j_1}v_{j'_2}
G^{{j'_3}{j'^*_1}{j^*_2}}_{{s^*}{j'^*_2}{j^*_1}}
G^{j'_3j_2^*{j'_1}^*}_{sj_1^*{j'_2}^*}\right)
\nonumber\\
&\quad\qquad\left( v_{j'_2}v_{j'_3}v_{j_2}v_{j_3}
G^{{j^*_1}{j'^*_2}{j'_3}}_{{s^*}{j_3}{j^*_2}}
G^{j_1^*{j'_3}{j'_2}^*}_{sj_2^*j_3}\right)
\end{align}
where the symmetry condition in \eqref{6jcond} was used in the
second equality.

Let us compare the formula in \eqref{SU2GSDReOrganized} with that
in \eqref{GSDsphere}. We set $j=j^*$ for all $j$ and drop all
stars, since all irreducible representations of $SU_k(2)$ are
self-dual. Then we find that the summation
\eqref{SU2GSDReOrganized} has the same form as the trace of
$D^{-1}\sum_{s}d_{s}\hat{B}_{p_2}^s\hat{B}_{p_3}^s\hat{B}_{p_1}^s$
on the graph on a sphere as in \eqref{GSDsphere},

\begin{align}
  \label{trBBB}
  &\text{tr}^{\text{torus}}(\frac{1}{D}\sum_{s}d_s\hat{B}_p^s)
  \nonumber\\
  =&\sum_{j_1j_2j_3}
  \Biggl\langle\bmm
  {\scalefont{0.6}
  \begin{tikzpicture}[scale=0.6]
  \draw [<-,>=latex] (2,0) arc (0:180:1 and 0.8);
  \draw [->,>=latex] (0,0) arc (180:360:1 and 0.8);
  \draw [->,>=latex] (2,0) -- (0,0);
  \node at(1,1.) {$j_1$};
  \node at(1,-0.6) {$j_2$};
  \node at(1,0.2) {$j_3$};
  \end{tikzpicture}
  }
  \emm\Biggr\vert
  \frac{1}{D}
  \sum_{s}d_{s}\hat{B}_{p_2}^s
  \hat{B}_{p_3}^s
  \hat{B}_{p_1}^s
    \Biggl\vert\bmm
  {\scalefont{0.6}
  \begin{tikzpicture}[scale=0.6]
  \draw [<-,>=latex] (2,0) arc (0:180:1 and 0.8);
  \draw [->,>=latex] (0,0) arc (180:360:1 and 0.8);
  \draw [->,>=latex] (2,0) -- (0,0);
  \node at(1,1.) {$j_1$};
  \node at(1,-0.6) {$j_2$};
  \node at(1,0.2) {$j_3$};
  \end{tikzpicture}
  }
  \emm\Biggr\rangle
  \nonumber\\
  =&\text{tr}^{\text{sphere}}(\frac{1}{D}\sum_{s}d_s\hat{B}_{p_2}^s
  \hat{B}_{p_3}^s\hat{B}_{p_1}^s)
\end{align}
where $\hat{B}_p^s$ is defined on the only plaquette $p$ on the
torus (see Fig. \ref{fig:reducedgraphs}(b)), while
$\hat{B}_{p_1}^s\hat{B}_{p_2}^s\hat{B}_{p_3}^s$ is defined on the
same graph on a sphere as in \eqref{GSDsphere} (see Fig.
\ref{fig:reducedgraphs}(a)).

The GSD on a torus becomes a trace on a sphere. The latter is
easer to deal with since the ground state on a sphere is
non-degenerate. The counting of ground states on a
torus turns into a problem dealing with excitations on the sphere.

In the following we evaluate the summation in the representation
of elementary excitations. let us introduce a new set of operators
$\{\hat{n}_p^r\}$ by a transformation,
\begin{align}
  \label{transformationnB}
  \hat{n}_p^r=\sum_{s}s_{r0}s_{rs}\hat{B}_p^s,
  \quad\hat{B}_p^s=\sum_{r}\frac{s_{rs}}{s_{r0}}\hat{n}_p^r
\end{align}
Here $s_{rs}$ is a symmetric matrix (referred to as the modular
$S$-matrix for $SU_k(2)$),
\begin{equation}
  \label{Smatrix}
  s_{rs}=\frac{1}{\sqrt{D}}\frac{\sin{\frac{(r+1)(s+1)\pi}{k+2}}}
  {\sin{\frac{\pi}{k+2}}}
\end{equation}
and has the properties 
\begin{align}
\label{Sproperty} s_{rs}=s_{sr},{\quad}s_{r0}=d_r/\sqrt{D}
\nonumber\\
\sum_{s}s_{rs}s_{st}=\delta_{rt}
\nonumber\\
\sum_{w}\frac{s_{wr}s_{ws}s_{wt}}{s_{w0}}=\delta_{rst}
\end{align}

Eq. \eqref{transformationnB} can be viewed as a finite discrete
Fourier transformation between $\{\hat{n}_p^r\}$ and
$\{\hat{B}_p^s\}$. By properties \eqref{Sproperty}, we see that
$\{\hat{n}_p^r\}$ are mutually orthonormal projectors, and they
form a resolution of the identity:
\begin{equation}
\hat{n}_p^r\hat{n}_p^s=\delta_{rs}\hat{n}_p^r, \quad
\sum_r{\hat{n}_p^r}=\text{id}
\end{equation}

In particular, $\hat{n}_p^0=\frac{1}{D}\sum_s{d_s}\hat{B}_p^s$ is
the operator $\hat{B}_p$ in the Hamiltonian. The operator
$\hat{n}_p^r$ projects onto the states with a quasiparticle
(labeled by $r$ type) occupying the plaquette $p$. Expressed as
common eigenvectors of $\{\hat{n}_p^r\}$, the elementary
excitations are classified  by the configuration of these
quasiparticles.

Particularly, on the graph on a sphere as in \eqref{trBBB}, the
Hilbert space has a basis of $\{\left|r_1,r_2,r_3\right\rangle\}$,
where only those $r_1$, $r_2$, and $r_3$ that satisfy
$\delta_{r_1r_2r_3}=1$ are allowed. Each basis vector
$\left|r_1,r_2,r_3\right\rangle$ is an elementary excitation with
the quasiparticles labeled by $r_1$, $r_2$, and $r_3$ occupying
the plaquettes $p_1$, $p_2$, and $p_3$. The configuration of
quasiparticles are globally constrained by
$\delta_{r_1r_2r_3}=1$\cite{HSW2}. Therefore, tracing opertors
$\{\hat{n}_p^r\}$ leads to
\begin{equation}
  \label{nnn}
  \text{tr}(\hat{n}_{p_2}^{r_2}\hat{n}_{p_3}^{r_3}
  \hat{n}_{p_1}^{r_1})=\delta_{r_2r_3r_1}
\end{equation}

Applying this rule reduces the summation \eqref{trBBB} to
\begin{align}
  &\text{tr} (\frac{1}{D}
  \sum_{s}d_{s}\hat{B}_{p_2}^s
  \hat{B}_{p_3}^s
  \hat{B}_{p_1}^s)
  \nonumber\\
  =&\text{tr}(
  \frac{1}{D}\sum_{s}d_s
  \sum_{r_1r_2r_3}
  \frac{s_{sr_1}s_{sr_2}s_{sr_3}}{s_{r_10}s_{r_20}s_{r_30}}
  \hat{n}_{p_2}^{r_2}\hat{n}_{p_3}^{r_3}\hat{n}_{p_1}^{r_1})
  \nonumber\\
  =&\sum_{r_1r_2r_3}\frac{1}{D}\sum_{s}d_s
  \frac{s_{sr_1}s_{sr_2}s_{sr_3}}{s_{r_10}s_{r_20}s_{r_30}}
  \delta_{r_1r_2r_3}
\end{align}
Then we substitute \eqref{dD}, \eqref{fusionrule} and \eqref{Smatrix} in 
and obtain  

\begin{align}
  \text{GSD}^{\text{torus}}_{SU_k(2)}
  =&\sum_{r_1,r_2,r_3=0}^{k}
  \frac{\sin{\frac{\pi}{k+2}}\delta_{r_1+r_2+r_3,2k}}
  {\sin{\frac{(r_1+1)\pi}{k+2}}\sin{\frac{(r_2+1)\pi}{k+2}}\sin{\frac{(r_3+1)\pi}{k+2}}}
  \nonumber\\
  =&\sum_{r=0}^{k}\sum_{s=0}^{r}\frac{\sin{\frac{\pi}{k+2}}}
  {\sin{\frac{(r+1)\pi}{k+2}}\sin{\frac{(s+1)\pi}{k+2}}\sin{\frac{(r-s+1)\pi}{k+2}}}
  \nonumber\\
  =&(k+1)^2  .
\end{align}
(Here we omit a rigorous proof of the last equality.)

We can also verify $\text{GSD}=(k+1)^2$ by a direct numerical
computation. We take the approach in \cite{MV} to
construct the numerical data of $6j$ symbols.
The construction depends on a parameter, the Kauffman
variable $A$ (in the same convention as in \cite{MV}),
which is specialized to roots of unity. We make the
following choice:
\begin{align}
\label{Achoice} \Biggl\{
\begin{array}{ll}
 A = \exp( \pi i /3)&\text{at }k=1\\
 A = \exp( 3 \pi i/8)&\text{at }k=2\\
 A = \exp( 3\pi i/5)&\text{at }k=3\\
\end{array}
\Biggr.
\end{align}

By this choice, the quantum dimensions $d_j$ take the values as
in \eqref{dD}, and the $6j$ symbols satisfy the self-consistent
conditions in \eqref{6jcond}. Using such data of quantum
dimensions $d_j$ and $6j$ symbols,
We compute the summation \eqref{torusGSD} at
\begin{align}
\label{NumericalGSD} \Biggl\{
\begin{array}{ll}
 \text{GSD} = 4  &\text{at }k=1\\
 \text{GSD} = 9  &\text{at }k=2\\
 \text{GSD} = 16 &\text{at }k=3\\
\end{array}
\Biggr.
\end{align}
which verifies $\text{GSD}=(k+1)^2$ in the particular cases.


\end{document}